\documentclass[prl,aps,superscriptaddress,showpacs]{revtex4}

\usepackage{graphicx}
\usepackage{epsfig}
\usepackage{dcolumn}
\usepackage{bm}
\usepackage{amsfonts,amsmath}
\usepackage{lineno}
\usepackage{setspace}

\newcommand{\ppbar}    {\mbox{${p\overline{p}}$}}

\newcommand{\jpsi}     {\mbox{$J/\psi$}}
\newcommand{\psiprime} {\mbox{$\psi(2S)$}}
\newcommand{\cchi}     {\mbox{$\chi_{c}$}}
\newcommand{\pt}       {\mbox{$p_{T}$}}
\newcommand{\pgev}     {\mbox{${\rm GeV}/c$}}
\newcommand{\mgev}     {\mbox{${\rm GeV}/c^2$}}

\newcommand{\ipb}      {\mbox{${\rm pb}^{-1}$}}

\newcommand{\cts}      {\mbox{${\rm cos}\;\theta^*$}}
\newcommand{\ccbar}    {\mbox{$c\overline{c}$}}
\newcommand{\Lxy}      {\mbox{$L_{xy}$}}
\newcommand{\ct}       {\mbox{$ct$}}
\newcommand{\fbkd}     {\mbox{$f_{bkd}$}}
\newcommand{\aeff}     {\mbox{$\alpha_{eff}$}}

\def\Journal#1#2#3#4{{#1} {\bf #2}, #3 (#4)}

\def \PRL      {Phys. Rev. Lett.}

\def \PRD      {Phys. Rev. D}

\def \PLB      {Phys. Lett. B}

\def \EPJC     {Eur. Phys. J. C}
\def \JPG      {J. Phys. G}
\def \etal     {\relax\ifmmode{et \; al.}\else{$et \; al.$}\fi}

\begin{document}
\title{Polarizations of $\jpsi$ and $\psiprime$ Mesons 
Produced in $\ppbar$ Collisions at $\sqrt{s}=1.96\;{\rm TeV}$}

\affiliation{Institute of Physics, Academia Sinica, Taipei, Taiwan 11529, Republic of China} 
\affiliation{Argonne National Laboratory, Argonne, Illinois 60439} 
\affiliation{Institut de Fisica d'Altes Energies, Universitat Autonoma de Barcelona, E-08193, Bellaterra (Barcelona), Spain} 
\affiliation{Baylor University, Waco, Texas  76798} 
\affiliation{Istituto Nazionale di Fisica Nucleare, University of Bologna, I-40127 Bologna, Italy} 
\affiliation{Brandeis University, Waltham, Massachusetts 02254} 
\affiliation{University of California, Davis, Davis, California  95616} 
\affiliation{University of California, Los Angeles, Los Angeles, California  90024} 
\affiliation{University of California, San Diego, La Jolla, California  92093} 
\affiliation{University of California, Santa Barbara, Santa Barbara, California 93106} 
\affiliation{Instituto de Fisica de Cantabria, CSIC-University of Cantabria, 39005 Santander, Spain} 
\affiliation{Carnegie Mellon University, Pittsburgh, PA  15213} 
\affiliation{Enrico Fermi Institute, University of Chicago, Chicago, Illinois 60637} 
\affiliation{Comenius University, 842 48 Bratislava, Slovakia; Institute of Experimental Physics, 040 01 Kosice, Slovakia} 
\affiliation{Joint Institute for Nuclear Research, RU-141980 Dubna, Russia} 
\affiliation{Duke University, Durham, North Carolina  27708} 
\affiliation{Fermi National Accelerator Laboratory, Batavia, Illinois 60510} 
\affiliation{University of Florida, Gainesville, Florida  32611} 
\affiliation{Laboratori Nazionali di Frascati, Istituto Nazionale di Fisica Nucleare, I-00044 Frascati, Italy} 
\affiliation{University of Geneva, CH-1211 Geneva 4, Switzerland} 
\affiliation{Glasgow University, Glasgow G12 8QQ, United Kingdom} 
\affiliation{Harvard University, Cambridge, Massachusetts 02138} 
\affiliation{Division of High Energy Physics, Department of Physics, University of Helsinki and Helsinki Institute of Physics, FIN-00014, Helsinki, Finland} 
\affiliation{University of Illinois, Urbana, Illinois 61801} 
\affiliation{The Johns Hopkins University, Baltimore, Maryland 21218} 
\affiliation{Institut f\"{u}r Experimentelle Kernphysik, Universit\"{a}t Karlsruhe, 76128 Karlsruhe, Germany} 
\affiliation{High Energy Accelerator Research Organization (KEK), Tsukuba, Ibaraki 305, Japan} 
\affiliation{Center for High Energy Physics: Kyungpook National University, Taegu 702-701, Korea; Seoul National University, Seoul 151-742, Korea; SungKyunKwan University, Suwon 440-746, Korea} 
\affiliation{Ernest Orlando Lawrence Berkeley National Laboratory, Berkeley, California 94720} 
\affiliation{University of Liverpool, Liverpool L69 7ZE, United Kingdom} 
\affiliation{University College London, London WC1E 6BT, United Kingdom} 
\affiliation{Centro de Investigaciones Energeticas Medioambientales y Tecnologicas, E-28040 Madrid, Spain} 
\affiliation{Massachusetts Institute of Technology, Cambridge, Massachusetts  02139} 
\affiliation{Institute of Particle Physics: McGill University, Montr\'{e}al, Canada H3A~2T8; and University of Toronto, Toronto, Canada M5S~1A7} 
\affiliation{University of Michigan, Ann Arbor, Michigan 48109} 
\affiliation{Michigan State University, East Lansing, Michigan  48824} 
\affiliation{University of New Mexico, Albuquerque, New Mexico 87131} 
\affiliation{Northwestern University, Evanston, Illinois  60208} 
\affiliation{The Ohio State University, Columbus, Ohio  43210} 
\affiliation{Okayama University, Okayama 700-8530, Japan} 
\affiliation{Osaka City University, Osaka 588, Japan} 
\affiliation{University of Oxford, Oxford OX1 3RH, United Kingdom} 
\affiliation{University of Padova, Istituto Nazionale di Fisica Nucleare, Sezione di Padova-Trento, I-35131 Padova, Italy} 
\affiliation{LPNHE, Universite Pierre et Marie Curie/IN2P3-CNRS, UMR7585, Paris, F-75252 France} 
\affiliation{University of Pennsylvania, Philadelphia, Pennsylvania 19104} 
\affiliation{Istituto Nazionale di Fisica Nucleare Pisa, Universities of Pisa, Siena and Scuola Normale Superiore, I-56127 Pisa, Italy} 
\affiliation{University of Pittsburgh, Pittsburgh, Pennsylvania 15260} 
\affiliation{Purdue University, West Lafayette, Indiana 47907} 
\affiliation{University of Rochester, Rochester, New York 14627} 
\affiliation{The Rockefeller University, New York, New York 10021} 
\affiliation{Istituto Nazionale di Fisica Nucleare, Sezione di Roma 1, University of Rome ``La Sapienza," I-00185 Roma, Italy} 
\affiliation{Rutgers University, Piscataway, New Jersey 08855} 
\affiliation{Texas A\&M University, College Station, Texas 77843} 
\affiliation{Istituto Nazionale di Fisica Nucleare, University of Trieste/\ Udine, Italy} 
\affiliation{University of Tsukuba, Tsukuba, Ibaraki 305, Japan} 
\affiliation{Tufts University, Medford, Massachusetts 02155} 
\affiliation{Waseda University, Tokyo 169, Japan} 
\affiliation{Wayne State University, Detroit, Michigan  48201} 
\affiliation{University of Wisconsin, Madison, Wisconsin 53706} 
\affiliation{Yale University, New Haven, Connecticut 06520} 
\author{A.~Abulencia}
\affiliation{University of Illinois, Urbana, Illinois 61801}
\author{J.~Adelman}
\affiliation{Enrico Fermi Institute, University of Chicago, Chicago, Illinois 60637}
\author{T.~Affolder}
\affiliation{University of California, Santa Barbara, Santa Barbara, California 93106}
\author{T.~Akimoto}
\affiliation{University of Tsukuba, Tsukuba, Ibaraki 305, Japan}
\author{M.G.~Albrow}
\affiliation{Fermi National Accelerator Laboratory, Batavia, Illinois 60510}
\author{S.~Amerio}
\affiliation{University of Padova, Istituto Nazionale di Fisica Nucleare, Sezione di Padova-Trento, I-35131 Padova, Italy}
\author{D.~Amidei}
\affiliation{University of Michigan, Ann Arbor, Michigan 48109}
\author{A.~Anastassov}
\affiliation{Rutgers University, Piscataway, New Jersey 08855}
\author{K.~Anikeev}
\affiliation{Fermi National Accelerator Laboratory, Batavia, Illinois 60510}
\author{A.~Annovi}
\affiliation{Laboratori Nazionali di Frascati, Istituto Nazionale di Fisica Nucleare, I-00044 Frascati, Italy}
% ...
\collaboration{CDF Collaboration\footnote{With visitors from $^a$University of Athens, 
$^b$University of Bristol, 
$^c$University Libre de Bruxelles, 
$^d$Cornell University, 
$^e$University of Cyprus, 
$^f$University of Dublin, 
$^g$University of Edinburgh, 
$^h$University of Heidelberg, 
$^i$Universidad Iberoamericana, 
$^j$University of Manchester, 
$^k$Nagasaki Institute of Applied Science, 
$^l$University de Oviedo, 
$^m$University of London, Queen Mary College, 
$^n$University of California Santa Cruz, 
$^o$Texas Tech University, 
$^p$University of California Irvine, and
$^q$IFIC(CSIC-Universitat de Valencia). 
}}
\noaffiliation


\begin{abstract}
We have measured the polarizations of $\jpsi$ and $\psiprime$ mesons 
as functions of their transverse momentum $\pt$ 
when they are produced promptly in the rapidity range 
$|y|<0.6$ with $\pt \geq 5 \;\pgev$. 
The analysis is performed using a data sample with an integrated 
luminosity of about $800\;\ipb$ collected by the CDF~II detector. 
For both vector mesons, we find that the polarizations become 
increasingly longitudinal as $\pt$ increases from $5$ to $30\;\pgev$. 
These results are compared to the predictions of nonrelativistic
quantum chromodynamics and other contemporary models.
The effective polarizations of $\jpsi$ and $\psiprime$ 
mesons from $B$-hadron decays are also reported.
\end{abstract}

\maketitle

An effective field theory, nonrelativistic quantum 
chromodynamics (NRQCD)~\cite{nrqcd}, provides a 
rigorous formalism for calculating the production 
rates of charmonium ($\ccbar$) states. NRQCD explains 
the direct production cross sections for $\jpsi$ and 
$\psiprime$ mesons observed at the Tevatron~\cite{run1-xs, run2-xs} 
and predicts their increasingly transverse polarizations
as $\pt$ increases, where $\pt$ is the meson's momentum
component perpendicular to the colliding beam 
direction~\cite{nrqcd-pol}. The first polarization 
measurements at the Tevatron~\cite{run1-pol} did not show such a trend. 
This Letter reports on $\jpsi$ and $\psiprime$ polarization measurements 
with a larger data sample than previously available. This allows 
the extension of the measurement to a higher $\pt$ region 
and makes a more stringent test of the NRQCD prediction. 

The NRQCD cross section calculation for $\ccbar$ production separates 
the long-distance nonperturbative contributions from the short-distance 
perturbative behavior. The former is treated as an expansion of the 
matrix elements in powers of the nonrelativistic charm-quark velocity. 
This expansion can be computed by lattice simulations, but currently the 
expansion coefficients are treated as universal parameters, which are adjusted 
to match the cross section measurements at the Tevatron~\cite{run1-xs, run2-xs}. 
The calculation also applies to $\ccbar$ production in $ep$ collisions, but 
HERA measurements of $\jpsi$ polarization tend to disagree with the NRQCD 
prediction~\cite{HERA}. These difficulties have led some authors to explore 
alternative power expansions of the long-distance interactions for the 
$\ccbar$ system~\cite{nrqcd-ext}. There are also new QCD-inspired models, 
the gluon tower model~\cite{khoze} and the $k_{T}$-factorization model~\cite{KT}, 
that accomodate vector-meson cross sections at both HERA and the Tevatron
and predict the vector-meson polarizations as functions of $\pt$.
These authors emphasize that measuring the vector-meson polarizations 
as functions of $\pt$ is a crucial test of NRQCD.

The CDF~II detector is described in detail elsewhere~\cite{run2-xs, CDF}.
In this analysis, the essential features are a muon system covering the 
central region of pseudorapidity, $|\eta|<0.6$, and the tracking system, 
immersed in the 1.4 T solenoidal magnetic field and composed of a 
silicon microstrip detector and a cylindrical drift chamber 
called the central outer tracker (COT). The data used here correspond 
to an integrated luminosity of about $800\;\ipb$ and were recorded 
between June 2004 and February 2006 by a dimuon trigger, which 
requires two opposite-charge muon candidates, each having  $\pt>1.5\;\pgev$. 

\begin{figure}[h]
\includegraphics[width=0.40\textwidth]{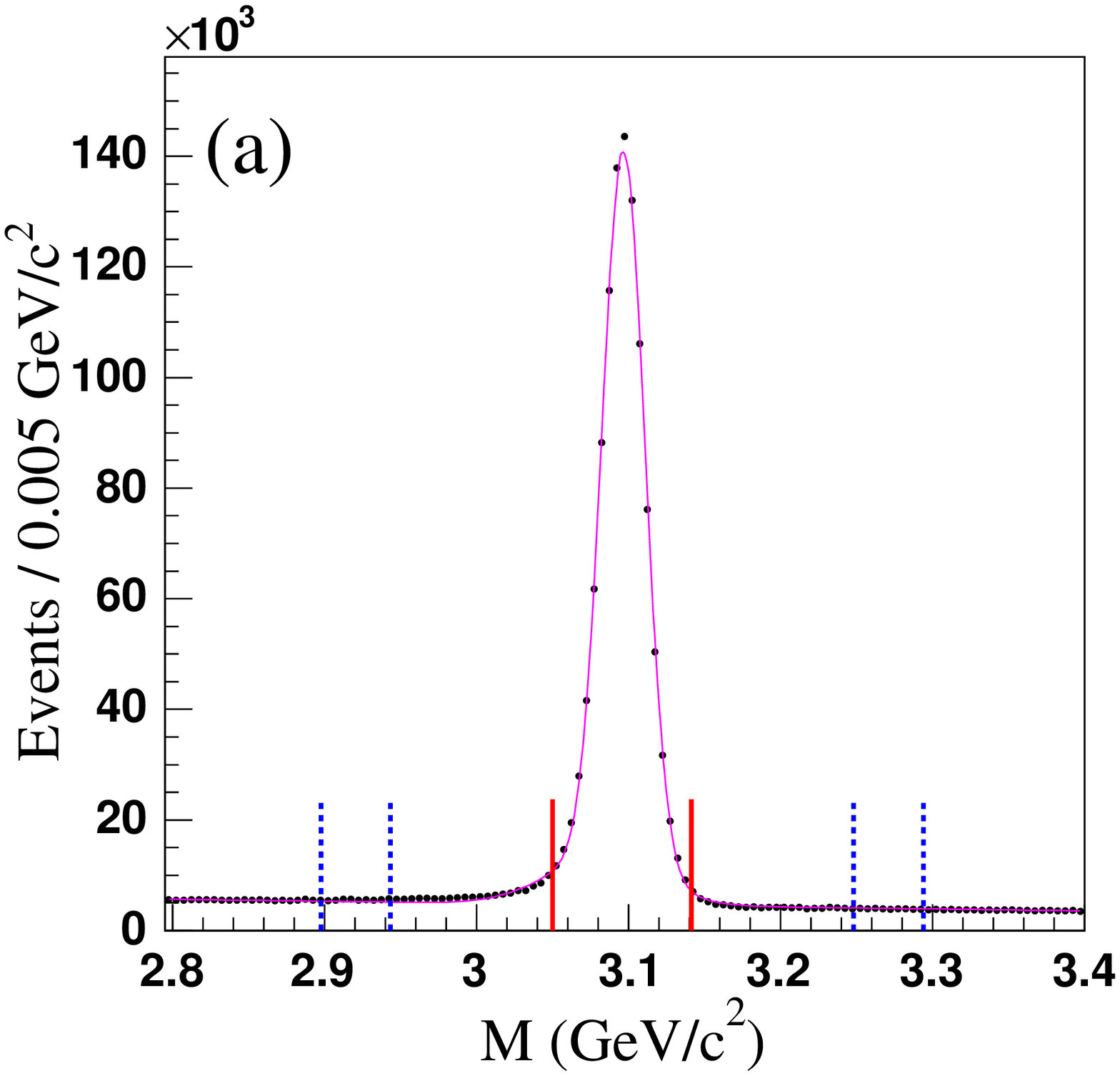}
\includegraphics[width=0.40\textwidth]{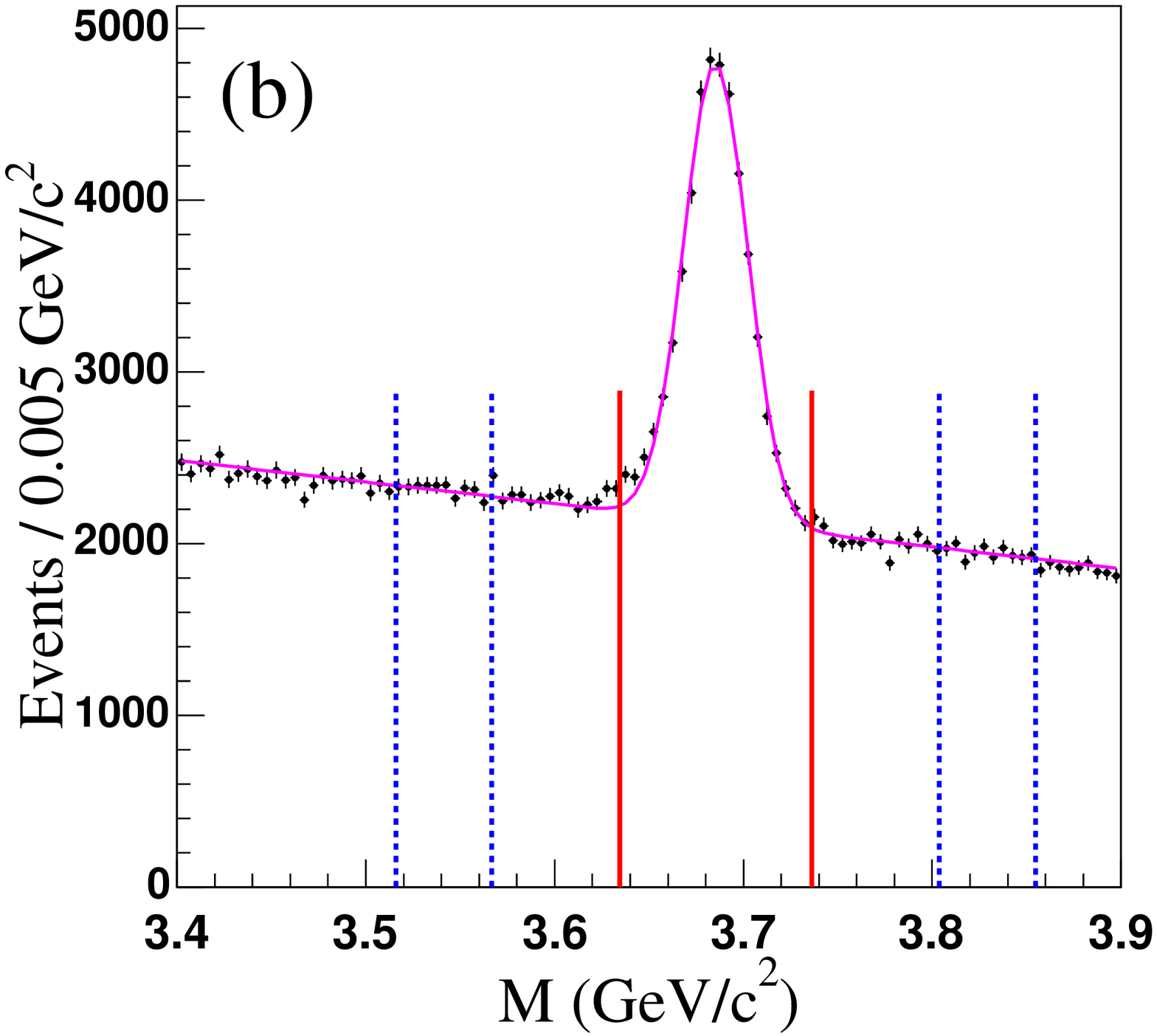}
\caption{Invariant mass distributions for (a) $\jpsi$ and (b) $\psiprime$ 
         candidates. The curves are fits to the data. The solid (dashed) 
         lines indicate the signal (sideband) regions.
         \label{mass}}
\end{figure}

Decays of 
vector mesons $V$ (either $\jpsi$ or $\psiprime$) $\rightarrow \mu^{+} \mu^{-}$ 
are selected from dimuon events for which each track has segments reconstructed 
in both the COT and the silicon microstrip detector. The $\pt$ of each muon is 
required to exceed $1.75\;\pgev$ in order to guarantee a well-measured trigger 
efficiency. The muon track pair is required to be consistent with originating 
from a common vertex and to have an invariant mass $M$ within the range 
$2.8\;(3.4) < M < 3.4\;(3.9)\;\mgev$ to be considered as a 
$\jpsi\;(\psiprime)$ candidate. To have a reasonable polarization sensitivity, 
the vector-meson candidates are required to have $\pt\geq5\;\pgev$ in the
rapidity range $|y\;(\equiv\frac{1}{2}{\rm ln}\frac{E+p_{||}}{E-p_{||}})|<0.6$,
where $E$ is the energy and $p_{||}$ is the momentum parallel to the beam 
direction of the dimuon system. Events are separated into a signal region 
and sideband regions, as indicated in Fig.~\ref{mass}. The fit to the data 
uses a double (single) Gaussian for the $\jpsi\;(\psiprime)$ signal and a 
linear background shape. The fits are used only to define signal and 
background regions. The signal regions are within $3\sigma_{V}$ of the 
fitted mass peaks $M_{V}$, where $\sigma_{V}$ is the width obtained in the 
fit to the invariant mass distribution. Both the background distribution and 
the quantity of background events under the signal peak are estimated by 
events from the lower and upper mass sidebands. The sideband regions are 
$7\sigma_{{\scriptsize\jpsi}}$ ($4\sigma_{{\scriptsize\psiprime}}$) away
from the signal region for $\jpsi$ ($\psiprime$).

\begin{figure}[h]
\includegraphics[width=0.40\textwidth]{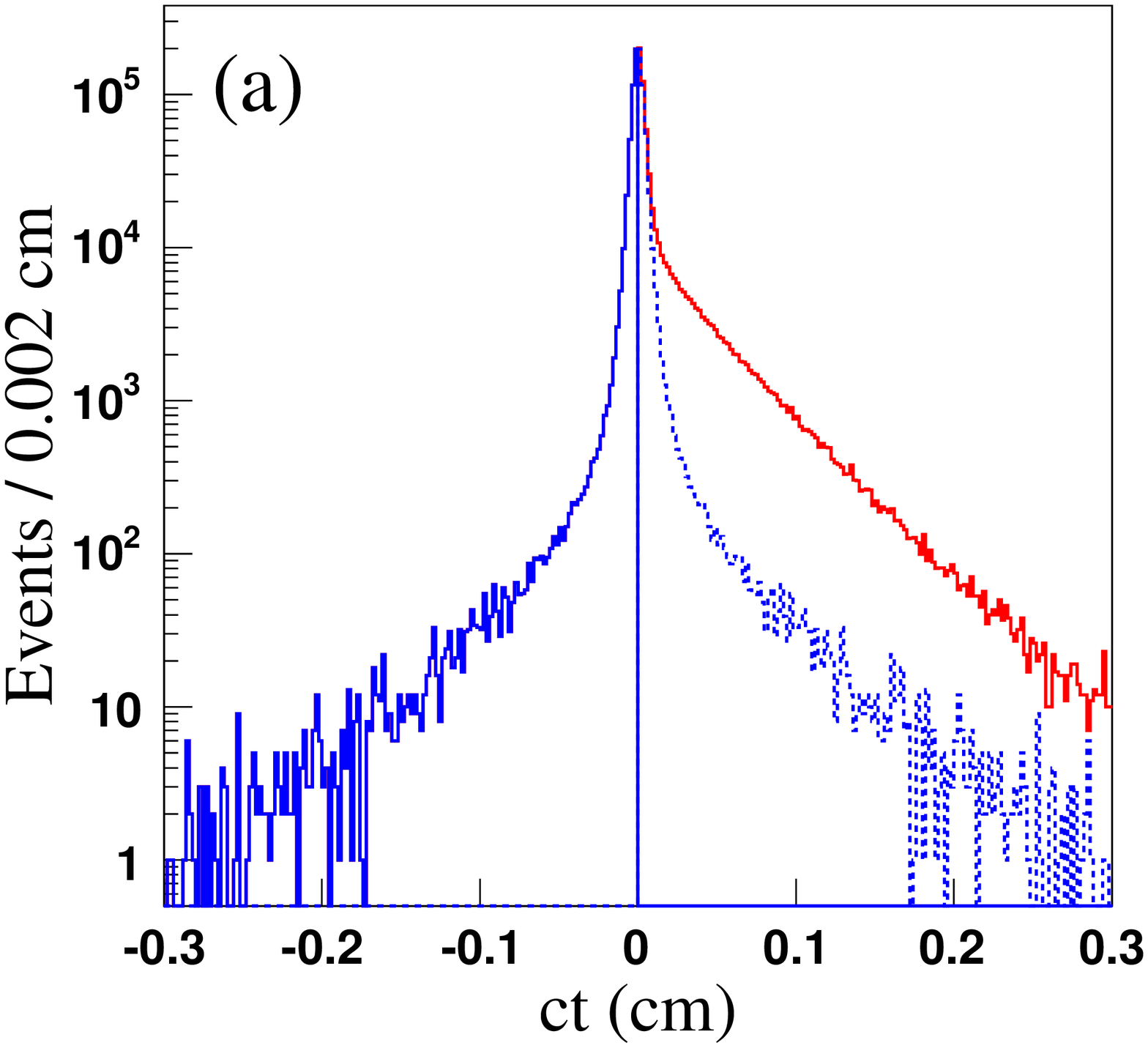}
\includegraphics[width=0.40\textwidth]{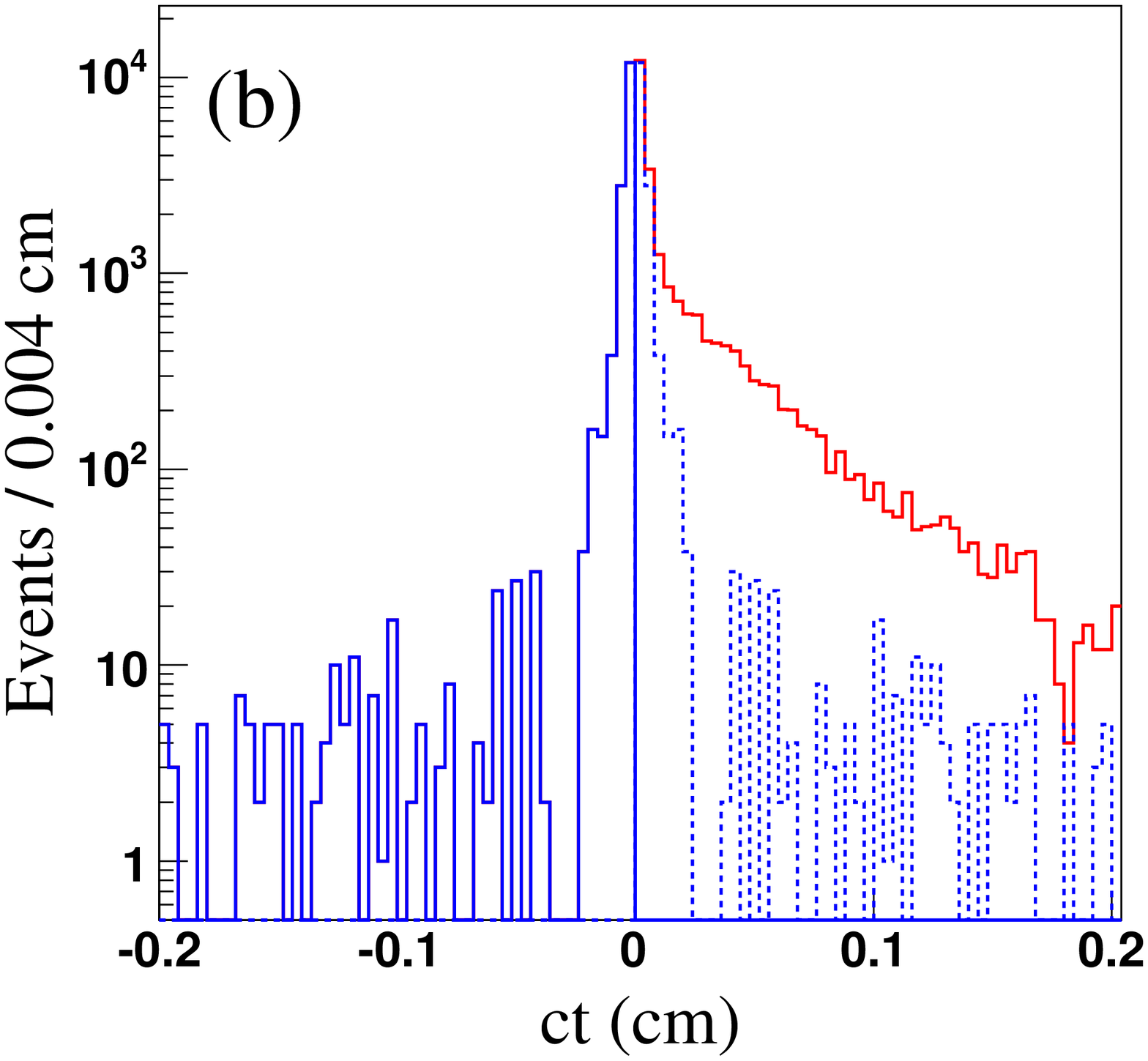}
\caption{Sideband-subtracted $\ct$ distributions for 
         (a) $\jpsi$ and (b) $\psiprime$ events. 
         The prompt Gaussian peak, positive excess from  
         $B$-hadron decays, and negative tail 
         from mismeasured events are shown. The dotted
         line is the reflection of the negative $ct$ 
         histogram about zero.
         \label{time}}
\end{figure}

For each candidate, we compute $\ct = M\Lxy/\pt$, where $t$ is the proper decay 
time and $\Lxy$ is the transverse distance between the beam line and the 
decay vertex in the plane normal to the beam direction.
The $\ct$ distributions of the selected dimuon events are shown in 
Fig.~\ref{time}. The $ct$ distribution of prompt events is a Gaussian distribution 
centered at zero due to finite tracking resolution. 
For $\jpsi$, the prompt events are due to direct production 
or the decays of heavier charmonium states such as $\cchi$ and $\psiprime$; 
for $\psiprime$, the prompt events are almost entirely due to direct production
since heavier charmonium states rarely decay to $\psiprime$~\cite{PDG}.
Both the $\jpsi$ and the $\psiprime$ samples contain significant numbers 
of events originating from long-lived $B$-hadron decays,  
as can be seen from the event excess at positive $\ct$.
We have measured the fraction of $B\rightarrow\jpsi+X$ events 
in the $\jpsi$ sample and found agreement with other results~\cite{run2-xs}. 
We select prompt events by requiring the
sum of the squared impact parameter significances of the 
positively and negatively charged muon tracks
$S\equiv(\frac{d_0^+}{\sigma^{+}})^2 + (\frac{d_0^-}{\sigma^{-}})^2\leq 8$.
The impact parameter $d_0$ is the distance of closest approach 
of the track to the beam line in the transverse plane. 
Vector-meson candidates from $B$-hadron decays are selected by requiring 
$S>16$ and $ct>0.03\;{\rm cm}$. This requirement retains a negligible 
fraction of prompt events in the $B$ sample.

To measure the polarizations of prompt $\jpsi$ 
and $\psiprime$ mesons as functions of $\pt$, the $\jpsi$ events are 
analyzed in six $\pt$ bins and the $\psiprime$ events in three bins, shown in
Table~\ref{tbl:pol}. 
We determine the fraction of $B$-decay 
background remaining in prompt samples $\fbkd$ by subtracting the number 
of negative $\ct$ events from the number of positive $\ct$ events.
Only a negligible fraction ($< 0.2\%$) of $B$ decays produce vector-meson 
events with negative $\ct$. For both vector mesons, $\fbkd$
increases with $\pt$, as listed in Table~\ref{tbl:pol}. The prompt 
polarization from the fitting algorithm is corrected for this contamination. 

The polarization information is contained in the distribution of the muon 
decay angle $\theta^{*}$, the angle of the $\mu^+$ in the rest frame of 
vector meson with respect to the vector-meson boost direction in the 
laboratory system. The decay angle distribution depends on the 
polarization parameter $\alpha$: 
$\dfrac{d\;N}{d\;\cts} \propto 1 + \alpha \;\rm{cos}^2 \theta^*\;(-1\le\alpha\le 1)$.  
For fully transverse (longitudinal) polarization, $\alpha = +1\;(-1)$.  
Intermediate values of $\alpha$ indicate a mixture of transverse 
and longitudinal polarization.

A template method is used to account for acceptance and efficiency. 
Two sets of $\cts$ distributions for fully polarized decays of $\jpsi$ 
and $\psiprime$ events, one longitudinal ($L$) and the other 
transverse ($T$), are produced with the CDF simulation program using the 
efficiency-corrected $\pt$ spectra measured from data~\cite{run2-xs, run2-psiprime-xs}.
We use the muon trigger efficiency measured using data as a function of track 
parameters ($\pt,\;\eta,\;\phi$) to account for detector non-uniformities. 
The parametrized efficiency is used as a filter on all simulated muons.
Events that pass reconstruction represent the behavior of fully polarized 
vector-meson decays in the detector.

\begin{figure}[h]
\includegraphics[width=0.50\textwidth]{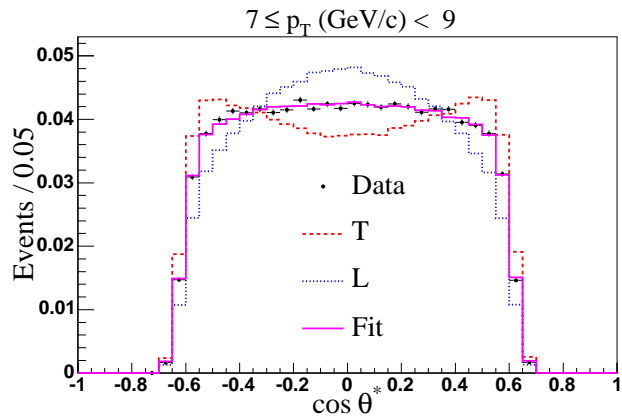}
\caption{$\cts$ distribution of data (points) and polarization fit for
the worst $\chi^2$ probability bin in the $\jpsi$ data. The 
dotted (dashed) line is the template for fully L (T) polarization. 
The fit describes the overall trend of the data well.
         \label{dist}}
\end{figure}

The fitting algorithm~\cite{run1-pol} uses two binned $\cts$ distributions 
for each $\pt$ bin, one made by $N_{S}$ events from the signal region 
(signal plus background) and the other made by $N_{B}$ events from the 
sideband regions (background). The $\chi^{2}$ minimization is done 
simultaneously for both $\cts$ distributions. 
The fitting algorithm 
includes an individual background term for each $\cts$ bin, normalized 
to $N_{B}$. Simulation shows that the $\cts$ resolution at all decay 
angles over the entire $\pt$ range is much smaller than the bin width 
of $0.05$ ($0.10$ for $\psiprime$) used here. 
The data, fit, and template distributions for the worst fit (9\% probability) 
in the $\jpsi$ data are shown in Fig.~\ref{dist}.  

All systematic uncertainties are much smaller than the statistical 
uncertainties.   
Varying the $\pt$ spectrum used in the simulation by $1\sigma$ 
changed the polarization parameter for $\jpsi$ 
at most by 0.002. 
A systematic uncertainty of 0.007 was estimated by the change in the 
polarization parameter when a modification was made on all trigger 
efficiencies by $\pm 1\sigma$. For $\psiprime$, the dominant systematic 
uncertainty came from the yield estimate because of the radiative tail 
and the large background. The total systematic uncertainties shown in 
Table~\ref{tbl:pol} were taken to be the quadrature sum of these 
individual uncertainties. Other possible sources of systematic 
uncertainties - signal definition and $\cts$ binning - were determined to be 
negligible. 
Corrections to prompt polarization from $B$-decay 
contamination were small, so that uncertainties on $B$-decay 
polarization measurements also had negligible effect.
No $\phi$-dependence of the polarizations was observed.

The polarization of $\jpsi$ mesons from inclusive $B_u$ and $B_d$ 
decays was measured by the {\it B{\footnotesize A}B{\footnotesize AR}} 
collaboration~\cite{BaBarB}. In this analysis, the $B$-hadron direction 
is unknown, so we define $\theta^{*}$ with respect to the $\jpsi$ 
direction in the laboratory system. The resulting polarization is 
somewhat diluted. As discussed in Ref.~\cite{run2-xs}, CDF uses a 
Monte Carlo procedure to adapt the {\it B{\footnotesize A}B{\footnotesize AR}}
measurement to predict the effective $\jpsi$ polarization parameter. 
For the $\jpsi$ events with $5\leq \pt < 30\;\pgev$, the CDF model for
$B_u$ and $B_d$ decays gives $\aeff = -0.145 \pm 0.009$, independent of $\pt$.
We have measured the polarization of vector mesons from $B$-hadron decays. 
For $\jpsi$, we find $\aeff=-0.106\pm 0.033 \;({\rm stat})\pm 0.007\;({\rm syst})$. 
At this level of accuracy, a polarization contribution by $\jpsi$ mesons 
from $B_{s}$ and $b$-baryon decays cannot be separated from the 
effective polarization due to those from $B_{u}$ and $B_{d}$ decays.
We also report the first measurement of the $\psiprime$ polarization 
from $B$-hadron decays: $\aeff=0.36 \pm 0.25 \;({\rm stat}) \pm 0.03\;({\rm syst})$.

\begin{figure}[ht]
 \begin{center}
  \includegraphics[width=0.45\textwidth]{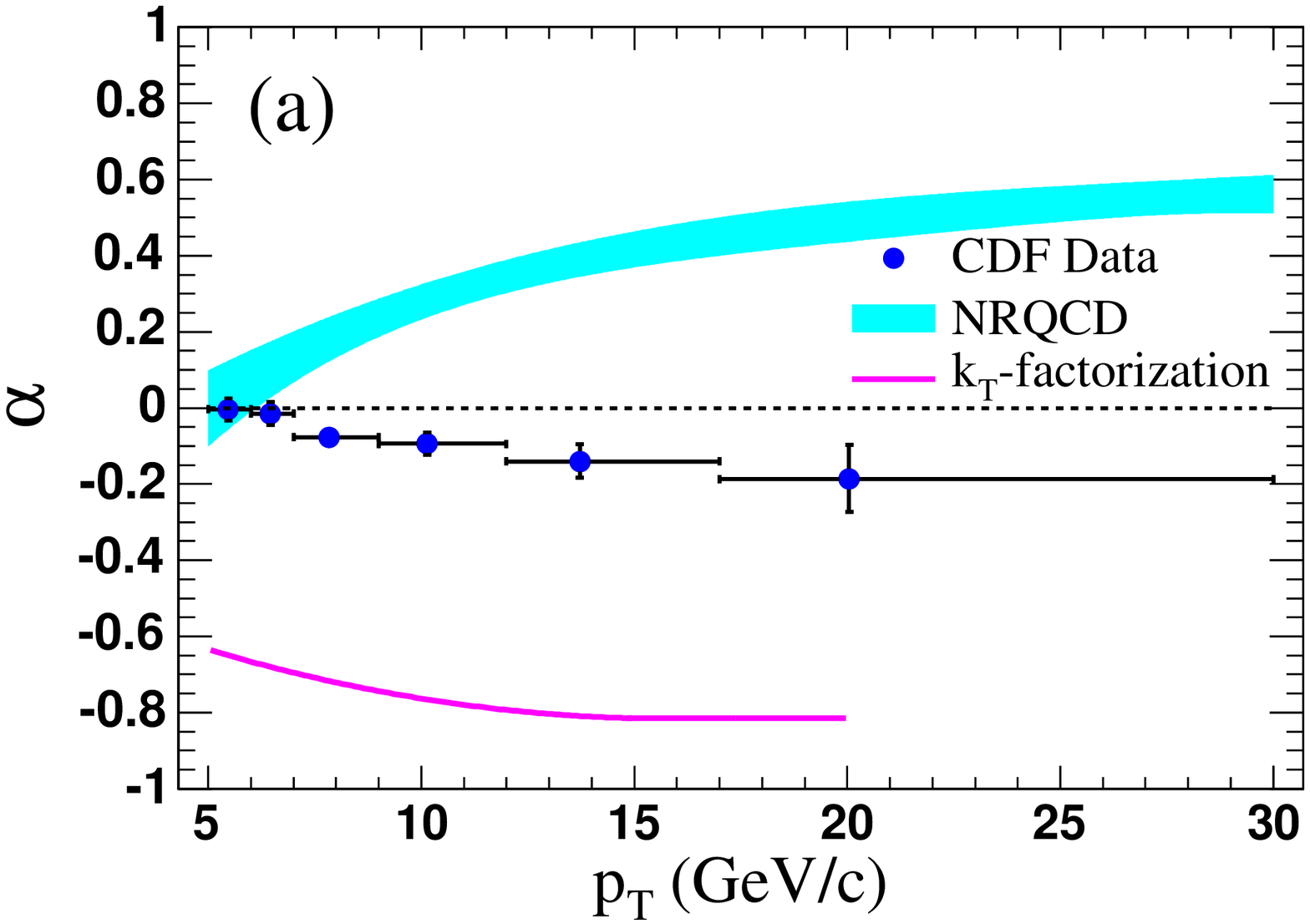}
  \includegraphics[width=0.45\textwidth]{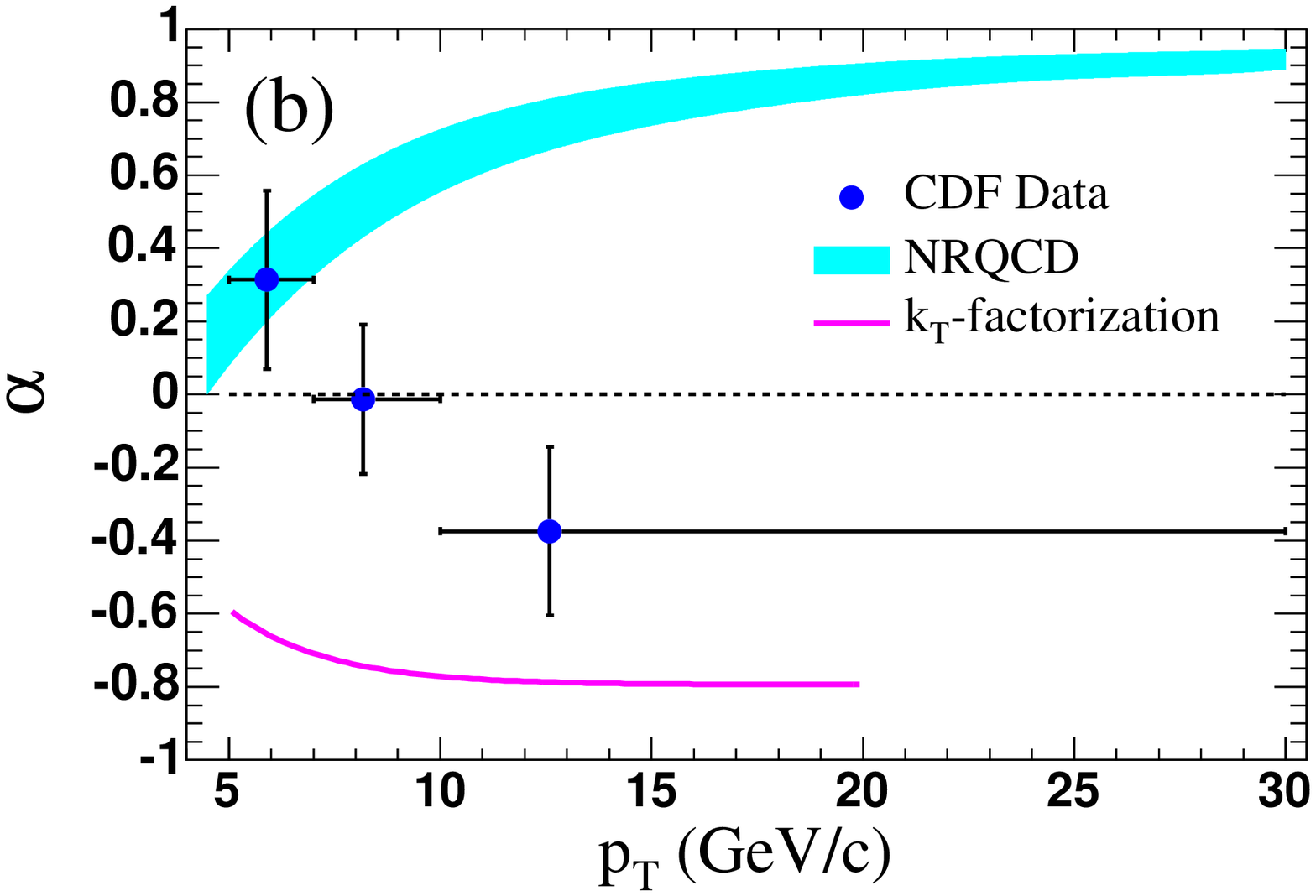}
  \caption{Prompt polarizations as functions of $p_T$: 
   (a) $\jpsi$ and (b) $\psiprime$. The band (line) is the prediction from 
   NRQCD~\cite{nrqcd-pol} (the $k_{T}$-factorization model~\cite{KT}).
   \label{fig:run2-polarization}}
 \end{center}
\end{figure}

The polarization parameters for both prompt vector mesons corrected 
for $\fbkd$ using our experimental results on $\aeff$ are listed as 
functions of $\pt$ in Table~\ref{tbl:pol} and are plotted in 
Fig.~\ref{fig:run2-polarization}. The polarization parameters for
$\jpsi$ are negative over the entire $\pt$ range of measurement and
become increasingly negative (favoring longitudinal polarization) 
as $\pt$ increases. For $\psiprime$, the central value of the 
polarization parameter is positive at small $\pt$, but, given the 
uncertainties, its behavior is consistent with the trend shown in 
the measurement of the $\jpsi$ polarization. 


The polarization behavior measured previously with $110\;\ipb$~\cite{run1-pol} 
is not consistent with the results presented here. This is a differential 
measurement, and the muon efficiencies in this analysis are true dimuon 
efficiencies. In Ref.~\cite{run1-pol}, they are the product of independent 
single muon efficiencies. The efficiency for muons with $\pt < 4 \;\pgev$ 
is crucial for good polarization sensitivity. In this analysis, the muon 
efficiency varies smoothly from $99\%$ to $97\%$ over this range. In the 
analysis of Ref.~\cite{run1-pol}, it varied from
$93\%$ to $40\%$ with significant jumps between individual data points.
Data from periods of drift chamber aging were omitted from this analysis
because the polarization results were inconsistent with the remainder of
the data. Studies such as this were not done in the analysis of Ref.~\cite{run1-pol}. 
The systematics of the polarization measurement are much better understood
in this analysis.

\begin{table*}[ht]
 \begin{center}
  \begin{tabular}{cccccc} \hline
   & $\pt {\small (\pgev)}$ & $<\!\pt\!> {\small (\pgev)}$ 
   & $\fbkd (\%)$ & $\alpha$ & $\chi^{2}/{\rm d.o.f}$  \\
   \hline \hline
   {\small $\jpsi$} & $5\!-\!6$   &  5.5 & $2.8\pm 0.2$ & $-0.004 \pm 0.029 \pm 0.009$ & 15.5/21 \\
                    & $6\!-\!7$   &  6.5 & $3.4\pm 0.2$ & $-0.015 \pm 0.028 \pm 0.010$ & 24.1/23 \\
                    & $7\!-\!9$   &  7.8 & $4.1\pm 0.2$ & $-0.077 \pm 0.023 \pm 0.013$ & 35.1/25 \\
                    & $9\!-\!12$  & 10.1 & $5.7\pm 0.3$ & $-0.094 \pm 0.028 \pm 0.007$ & 34.0/29 \\
                    & $12\!-\!17$ & 13.7 & $6.7\pm 0.6$ & $-0.140 \pm 0.043 \pm 0.007$ & 35.0/31 \\
                    & $17\!-\!30$ & 20.0 & $13.6\pm 1.4$& $-0.187 \pm 0.090 \pm 0.007$ & 33.9/35 \\ 
   \hline\hline
   {\small $\psiprime$} & $5\!-\!7$   & 5.9   & $1.6\pm 0.9$ & $+0.314\pm 0.242\pm 0.028$  & 13.1/11 \\
                        & $7\!-\!10$  & 8.2   & $4.9\pm 1.2$ & $-0.013\pm 0.201\pm 0.035$  & 18.5/13 \\
                        & $10\!-\!30$ & 12.6  & $8.6\pm 1.8$ & $-0.374\pm 0.222\pm 0.062$  & 26.9/17 \\
   \hline
   \end{tabular}
  \caption{Polarization parameter $\alpha$ for prompt production in each $p_T$ bin. 
           The first (second) uncertainty is statistical (systematic).
           $<\!\pt\!>$ is the average transverse momentum. 
           \label{tbl:pol}}
 \end{center}
\end{table*}

These polarization measurements for the charmed vector mesons extend to a 
$\pt$ regime where perturbative QCD should be applicable. The results are 
compared to the predictions of NRQCD and the $k_{T}$-factorization model
in Fig.~\ref{fig:run2-polarization}. The prediction of the 
$k_{T}$-factorization model is presented for $\pt<20\;\pgev$ and does not 
include the contribution from the decays of heavier charmonium states for 
$\jpsi$ production. The polarizations for prompt production of both vector 
mesons become increasingly longitudinal as $\pt$ increases beyond $10\;\pgev$. 
This behavior is in strong disagreement with the NRQCD prediction of large 
transverse polarization at high $\pt$. 
It is striking that the NRQCD calculation and the other models reproduce 
the measured $\jpsi$ and $\psiprime$ cross sections at the Tevatron, but fail
to describe the polarization at high $\pt$.
This indicates that there is some important aspect of the production mechanism that is 
not yet understood. 

We thank the Fermilab staff and the technical staffs of the participating 
institutions for their vital contributions. This work was supported by the 
U.S. Department of Energy and National Science Foundation; the Italian 
Istituto Nazionale di Fisica Nucleare; the Ministry of Education, Culture, 
Sports, Science and Technology of Japan; the Natural Sciences and 
Engineering Research Council of Canada; the National Science Council 
of the Republic of China; the Swiss National Science Foundation; the 
A.P. Sloan Foundation; the Bundesministerium f{\"u}r Bildung und Forschung, 
Germany; the Korean Science and Engineering Foundation and the Korean Research 
Foundation; the Particle Physics and Astronomy Research Council and the Royal 
Society, UK; the Institut National de Physique Nucleaire et Physique des Particules/CNRS; 
the Russian Foundation for Basic Research; the Comisi{\'o}n 
Interministerial de Ciencia y Tecnolog\'{\i}a, Spain; the European Community's 
Human Potential Programme; the Slovak R\&D Agency; and the Academy of Finland.

\clearpage



\begin{thebibliography}{99}
\bibitem{CDF}
  The CDF coordinate system has $\hat{z}$ along the proton direction, 
  $\hat{x}$ horizontal pointing outward from the Tevatron ring, and 
  $\hat{y}$ vertical. $\theta$ ($\phi$) is the polar (azimuthal) angle 
  measured with respect to $\hat{z}$, and $\eta$ is the pseudorapidity 
  defined as $-{\rm ln}\;({\rm tan}\;(\theta/2))$. The transverse 
  momentum of a particle is denoted as $\pt=p\;{\rm sin}\;\theta$.
\end{thebibliography}
\end{document}